\documentclass[10pt,conference, letterpaper]{IEEEtran}
\usepackage{float}
\usepackage{graphicx}
\usepackage{caption}
\usepackage{subcaption}
\usepackage{parskip}
\usepackage{mathtools}
\usepackage{enumitem}
\usepackage{algorithm}
\usepackage{algorithmic}
\usepackage{textcomp}
\usepackage{lscape}
\usepackage{longtable}  
\usepackage{amsmath}
\usepackage{wrapfig}
\usepackage{blindtext}
\usepackage{dblfloatfix}
\graphicspath{{Pictures/}}
\usepackage{color}
\usepackage{multirow}
\usepackage{array}
\usepackage{cite}
\usepackage[latin1]{inputenc}
\usepackage{tikz}
\usetikzlibrary{shapes,arrows}
 \IEEEoverridecommandlockouts        

\def\BibTeX{{\rmfamily B\kern-.05em{\scshape i\kern-.025em b}\kern-.08em \TeX}}

\ifCLASSINFOpdf
\else
\fi
\begin{document}
\title{Multi-level Dynamic Optimization of  Intelligent LEACH with Cost Effective Deep Belief Network}

\author{Muhammad U. Javed, Zaid Bin Tariq, Usama Muneeb, Ijaz Haider Naqvi
\thanks{All  authors were affiliated with the Lahore University of Management Sciences during this work. (e-mail:
	17100136@lums.edu.pk;  zaid.tariq@lums.edu.pk;
	17100273@lums.edu.pk; ijaz.naqvi@lums.edu.pk)}}
\maketitle
\begin{abstract}

 Energy utilization is a key attribute for energy constrained wireless sensor networks (WSN) that directly impacts the life time of the network.  LEACH (and its variants) are considered to be the most common energy efficient routing protocols for WSN. In this paper, we propose an optimized modification of LEACH that makes use of multi-hop communication, dynamic cluster boundaries and energy conservation in routing to maximize lifetime of a network. We propose a multi-level approach to maximize our gains with regards to energy conservation i.e., i) \emph{Dynamic programming} based intra-cluster optimization technique has been proposed ii) \emph{Ant Colony Optimization} is used for energy efficient cluster head connection with sink node and iii) \emph{Voronoi Tessellation} are employed for efficient coverage planning i.e., dynamic formation of cluster boundaries. In order to accommodate a more flexible adhoc network, hybrid (reactive and proactive) event monitoring based on \emph{Deep Belief Network} has been integrated in distributed nodes to improve the latency of the system. The results show that the proposed scheme significantly outperforms the current state of the art with regards to network lifetime and throughput.
\end{abstract}

\IEEEpeerreviewmaketitle
\section{Introduction}

Wireless sensor networks (WSN) consist of autonomous nodes for monitoring different attributes of an environment. The basic functions of WSNs include sensing, collecting, processing and distributing information. WSNs typically consist of static nodes that need to continue working for a long period of time. Thus, a large number of  researchers focus on the energy consumption and network life time of a WSN.  The sensor nodes are generally battery-powered devices and thus energy efficient routing directly translates into a larger life time of a network. In recent years, four different classes of routing protocols have been evolved with numerous routing protocols under each category. These classes are  geographical  routing protocols, data centric routing protocols, clustering-based routing protocols and hybrid routing protocol [\ref{leachEEE14}]. Amongst these categories, clustering based hierarchical routing protocols have been very popular because of their scalability. The clustering based routing protocols divide the sensor nodes into groups which are connected to a local Base Station (BS) or a Cluster head (CH). A large number of clustering based routing protocols have been proposed in the literature such as LEACH, LEACH-C, KLEACH, s-LEACH, EEE-LEACH etc. [\ref{leachEEE14}]. Amongst them, LEACH is the first and most popular clustering based hierarchical routing protocol for WSNs. LEACH (Low-Energy Adaptive Clustering hierarchy) is a self-organizing, adaptive clustering protocol that uses randomization based probability to distribute the energy load equally to the sensor nodes in the network. In LEACH protocol, the nodes have the ability to organize themselves into clusters with one node acting as a CH (or router) which aggregates data for other nodes. This makes randomized rotation of the high energy nodes as cluster heads and thus conserves energy (or battery drain) of all sensor nodes resulting in an increased network lifetime [\ref{leachEEE5}]. LEACH also performs the data aggregation and data fusion (data compression)  at cluster head level before transmitting data to sink node, further reducing the energy consumption and enhancing the network lifetime with application specific data processing.

There are two phases of LEACH protocol; i) the \emph{setup phase} which includes Cluster Head (CH) selection, cluster setup phase and cluster scheduling and ii)  2) the \emph{steady phase} which includes data aggregation, compression and transmission. LEACH is however unstable during setup phase because of its dependence on the density of the sensor nodes. As multi-hop communication is not employed, large networks consume significant energy in transmission of data from nodes farther from the sink node.  The main factors affecting the energy consumption are: perceptual data, data processing and most importantly radio (RF) communication. In this paper, instead of using three levels of routing as in [\ref{leachEEE14}], [\ref{leachEEE5}], we propose two levels of clustering  and make use of dynamic optimization and stochastic geometry techniques to improve the energy consumption of the network. In the following, we list the key contributions of our work:

\begin{enumerate}
\item \textbf{Maximum Network Lifetime:}
Multi-hop intra-cluster event propagation and hybrid (multi-hop and parallel) event propagation at inter cluster level have been proposed which considerably improve the network lifetime.

\item \textbf{Coverage Planning: }
A dynamic coverage range for clusters has been proposed; the coverage range of clusters and their boundaries are determined using Voronoi Tessellation.

\item \textbf{Latency-Flexible Monitoring:}
Latency or network delay is the time elapsed between the measurement reading and the transmission of data to the sink node. The proposed multi-hop routing approach would impact the latency of the system. In order to compensate that, we have proposed \emph{Deep Belief Network (DBN)} based event monitoring which switches between proactive and reactive event monitoring. DBN network reduces the dimensionality of the input data and  allows faster processing resulting in an improved latency of the system. 
   
\item \textbf{Distributed Multi agent Intelligence (DMI):}
A distributed network does not have a single point of failure. We design a reliable and robust system that does not fail by the death of a particular sensor node in the field by dynamic optimization at intra-cluster level. We assume that the sink node is located outside of the field in a secure environment with large energy supply.

\end{enumerate}

\section{Prior Art and Our Work}
\subsection{Routing Framework}
Graph theory based optimization has been used for routing in wireless sensor networks [\ref{leachEEE1}]-[\ref{leachEEE6}]. The key objective of graph theoretic approaches is to minimize the overall power consumption in a Wireless Sensor Network (WSN)  [\ref{leachEEE16}, \ref{leachEEE17}]. In [\ref{leachEEE1}], the distributed tree based approach is used for distributed collision handling but this approach results in a switching delay and most of the network remains passive. Therefore, a dynamic approach is required to tackle this problem. Within the dynamic optimization frameworks, the network agility is enhanced by exposing low level details of packet collisions [\ref{leachEEE3}].  In [\ref{leachEEE2}], \emph{Ant Colony Optimization} (ACO) has been used for routing where all routing nodes try to establish a parallel path to the sink node. 

In our proposed framework, rather than connecting all the cluster heads in parallel to base station, we have used ACO for minimum cost routing [\ref{leachEEE13}] between cluster heads and the sink node. Furthermore, since energy minimization is the foremost concern in WSNs, proximity based Voronoi tessellations have been proposed (rather than hierarchical broadcasting) so as to keep track of the broadcast and multihop links and provide energy optimization with maximum terrain coverage. The intra-cluster routing is locally distributed and is based on parallel computation [\ref{leachEEE4}] to form a connected dominating set (CDS) [\ref{leachEEE6}] which does not require separate query handling. In order to incorporate coverage planning, dynamic cluster boundaries using Voronoi Tessellations have been used which are centrally controlled at the sink node  [\ref{leachEEE7}]-[\ref{leachEEE15}]. The comparative analysis of network lifetime and energy efficiency has been carried out in [\ref{leachEEE5}] and on that basis, we have selected LEACH EEE as a comparative measure.

\subsection{Assumptions of Network Model}

The network consists of large number of randomly distributed nodes. The area is divided into clusters based on node density. Each cluster has a Cluster Head (CH). In our implementation, we do not account for message retransmissions caused due to collision, as we are interested in studying the effective improvement caused by each algorithm. The following assumptions are made in our network model:
\begin{enumerate}[label=(\roman*)]
	\item  We have a strongly connected network
	
	\item Each node with distinct ID would transmit/receive messages along its adjacent links/edges
 
 \item Each node is aware if it is a CH or not 
 
 \item Each node in the network knows the IDs of it's neighbors in the graph 
 
	\item Each node can transmit and receive data packets 

\end{enumerate}
\section{Proposed Solution}
 We propose a hierarchical network which is divided into cluster regions by implementing LEACH node proximity algorithm. The designed protocol generates an echo based event request between connected motes and sink node by two way communication. It also provides information about the location of event due to its geometry of strongly connected motes. Then in order to minimize the cost of broadcast messages at cluster level (CH to CH), \emph{ACO}  is implemented which identifies the path to sink node with minimum cost during inter cluster multi-hop communication. Within each cluster, a random CH is allocated on the basis of the cost function which helps in minimizing the overall energy resources for a durable network. 

We propose a modified LEACH algorithm which minimizes the overall energy consumption of the network and at the same time achieves maximum flow of the network. In order to achieve our goals, we introduce dynamic cluster boundaries; the sink node broadcasts beacon signals to every node in the network in order to come up with a distance vector based coverage. This allows the sink node to have an idea about the network and helps it to form optimized cluster boundaries using Voronoi tessellations. Thus, dynamic cluster boundaries are formed on the basis of node density and energy consumptions. Voronoi Tessellations allow us to minimize energy and maximize the radius of each cluster. Note that in LEACH, the CH is selected randomly after a certain period of time. So whenever a new CH is randomly selected, Voronoi Tessellation is used again at the sink node to reformulate the cluster boundaries. Veronoi Tessellation algorithm is run at the sink node from the time of new CH formation up to the time of convergence of network. Once the cluster boundaries have been finalized, ACO, which runs at the sink node, is used to determine the path with minimum cost from the CHs to the sink in a multi-hop communication model.

\tikzstyle{decision} = [diamond, draw, fill=blue!20, 
    text width=4.5em, text badly centered, node distance=3cm, inner sep=0pt]
\tikzstyle{block} = [rectangle, draw, fill=blue!20, 
    text width=10em, text centered, rounded corners, minimum height=2em]
\tikzstyle{line} = [draw, -latex']
\tikzstyle{cloud} = [draw, ellipse,fill=red!20, node distance=3cm,
    minimum height=2em]
    
		\begin{figure}
		\centering
\begin{tikzpicture}[node distance = 2cm, auto]
    \node [block] (init) {Start};
    \node [block, below of=init] (initialdistance) {Base Station (BS) Broadcasts signals to nodes and approximates distance accordingly};
    \node [block, below of=initialdistance] (initialclusters) {BS runs Voronoi Tesselation for cluster formation and generate random cluster heads};
    \node [block, below of=initialclusters] (randomnumber) {Periodically (after every $T$ seconds), generates a random number};
    \node [decision, below of=randomnumber] (decide) {Less than a threshold ?};
    \node [block, below of=decide, node distance=3cm] (start) {Start campaign for cluster head};
		\node [block, right of=decide, node distance=4cm] (wait) {Wait for the next campaign};
		\node [block, below of=start] (clusterhead) {Select node with largest energy reserve as the cluster head (CH)};                                                   
  \node[block, below of=clusterhead](voronoi) {Re-Run Voronoi Tesselation for optimal cluster boundaries};
 \node[block, below of=voronoi](ACO){CH uses ACO to find best path to base station};
 \node[block, below of=ACO](end){Cluster sends data to base station};
    \path [line] (init) -- (initialdistance);
    \path [line] (initialdistance) -- (initialclusters);
    \path [line] (initialclusters) -- (randomnumber);
		\path [line] (randomnumber) -- (decide);
    \path [line] (decide) -- node {Yes} (start);
		\path [line] (decide) -- node {No} (wait);
		\path [line, dashed] (wait) |- (randomnumber);
		\path [line] (start) -- (clusterhead);
    \path [line] (clusterhead) -- (voronoi);
    \path [line] (voronoi) -- (ACO);
    \path [line] (ACO) -- (end);

\end{tikzpicture}
	\caption{High level working of the network} \label{flowchart}
\end{figure}
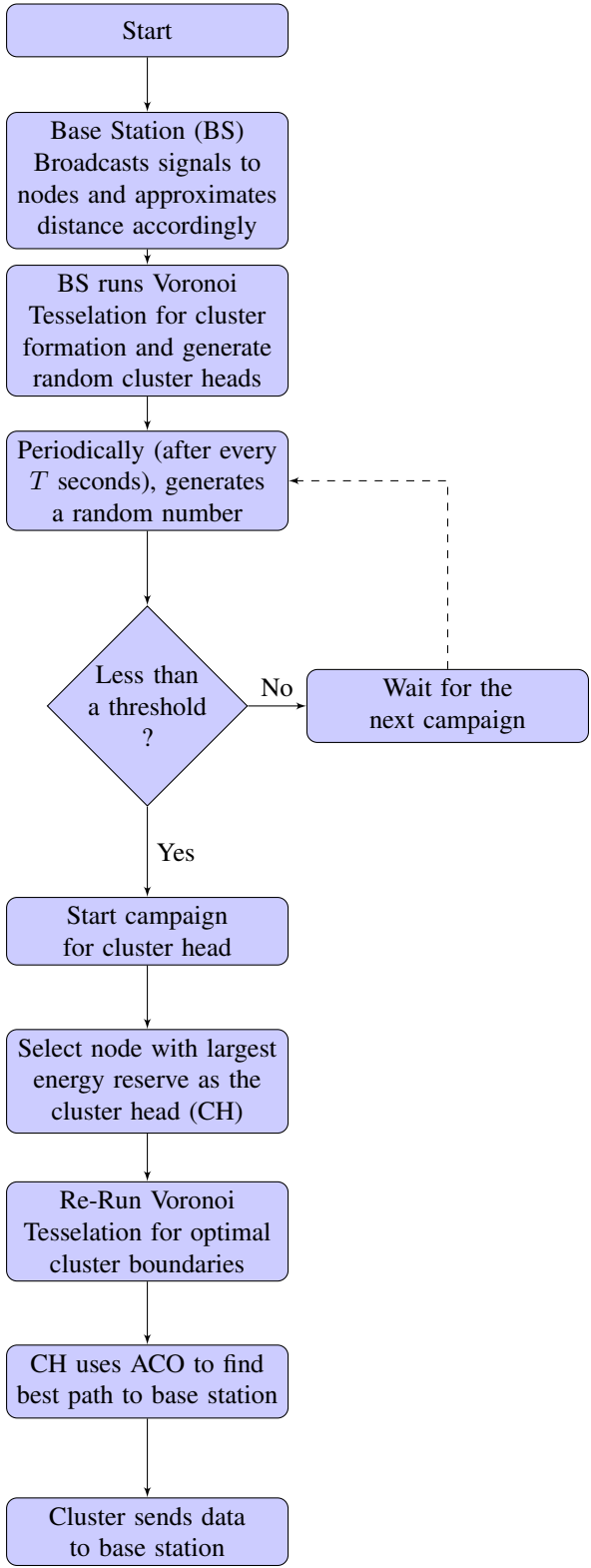

Figure \ref{flowchart} summarizes the high level working of the network. Sink node broadcasts a message to all nodes in the network and receives back the beacon signals. It approximates the distance from the sensing node on the basis of the received signal strength indicator (RSSI). The sink node then forms clusters of sensing nodes with similar RSSI values and determines the initial cluster boundaries using Voronoi Tessellation algorithm such that the radii are maximized. In each cluster, a CH is selected randomly in the start-up phase. Thereafter, the sink node generates a random number periodically for cluster head re-selection. If the number is less than a stochastic threshold, then on the basis of intra cluster available energy resources, CHs are re-selected. The sink node informs the the new CHs through a broadcast message. After cluster head selection, Voronoi tessellations are updated at sink node while keeping energy constant and maximizing cluster radii. If the generated number turns out to be less than the threshold, the sink node waits for the next campaign for cluster head re-selection.  Thereafter, in order to conserve the energy required for transmissions from CHs to the sink node, ant colony optimization (ACO) is used to find the best path from different CHs to the sink node. Thus, the network topology consists of two hierarchical levels: i) centralized CH-Sink node hierarchy ii) distributed intra-cluster hierarchy.  
  
\subsection{Centralized CH-Sink Node Hierarchy}  
The inter cluster network topology in centralized. The sink node decides with regards to the cluster heads, cluster boundaries as well as about the routing between different clusters. In the following we explain all these one by one.
\subsubsection{Selection of Cluster Head}
Once the campaign for the cluster head has started, the sink node finds out the node with the highest energy reserve with the help of the following function [\ref{leachEEE5}]:

\begin{equation}
\label{thresh}
T(n) = \frac{p}{1-p(r\mod(1/p))}\frac{E_{res}}{E_o} 
\end{equation}
where $1/p$ is the total number of nodes in a cluster, $r$ represents the number of time CHs are being reselected, $E_{res}$ is the residual energy of the node and $E_o$ is the initial energy of the node. The value of the above mentioned function is computed for all the nodes in the cluster and the node with the largest $T(n)$ is selected as the cluster head. Once new cluster heads have been selected, the network is informed about the new set of CHs through a broadcast message.

\subsubsection{Dynamic Cluster Boundaries}
For a given network with $d_{max}$ and $d_{min}$ be the maximum and minimum inter node distances respectively, the cluster radius is evaluated based on distance between each node and sink node (SN) using the following equation
 \begin{equation}
 \label{veronoi}
 R_c= 1-\alpha \frac{d_{max}-d(n_{i},SN)}{d_{max}-d_{min}}  
\end{equation}
where $R_c$ is the radius of the $c^{th}$ cluster where $c\in\{ 1,2,...,C\}$ with $C$ be the total number of clusters in the network, $\alpha$ indicates normalized energy per cluster radius and its value lies between 0 and 1. The algorithm converges when $\alpha$ $\rightarrow$ $\frac{\sum\limits_{i=1}^n E_{i}}{R_{i}}$.
\subsection{Ant Colony Optimization }
Ant Colony Optimization (ACO) is a swarm based optimization technique that is used to solve combinatorial optimization problems efficiently. ACO is inspired from the food collecting patterns of an ant colony in such a way that an ant leaves trail of pheromone hormones on its path of visit to it's colleagues in identifying the objective (food). At each point the ant has two options, either to explore the objective heuristically or leverage the pheromone trails left by other colleagues. A set of these probabilistic rules at each point, helps to minimize the total cost of visits and optimally find an optimum path through multiple points. Let an ant be a short term memory based stochastic path allocator. At each step,  in order to find the minimum path from CH to sink node, an ant applies random proportional rule to decide which next cluster to visit. The probability with which an ant currently at cluster $i$ chooses to go to cluster $j$ is given by the following equation:
\begin{equation}
\label{prob}
P_{ij}= \gamma \tau_{ij}^{\alpha}\eta_{ij}^{\beta}
\end{equation}
where $\gamma $ is a normalization constant ranging between 0 and 1,  $\tau$ indicates pheromone trail, $\eta$ indicates heuristic value that is available apriori and finally $\alpha$ and $\beta $ determine the relative influence of pheromone trail and heuristic information. If $\alpha $=0, closer clusters are more likely to be selected which suggests a stochastic greedy algorithm. If $\beta$=0, only pheromone is amplified and heuristic approach is not used. The heuristic value for two adjacent cluster heads $i$ and $j$ is defined by the following equation 
\begin{equation}
\label{eta}
\eta_{ij}= \Delta\left(1+ \left(\frac{d(j,s)}{d(i,s)}\right)^\lambda\right)
\end{equation} 
where $\Delta = \frac{E_i}{d(i,j)}$ represents the energy utilization from $i$ to $j$ and is normalized through $i$ to $j$ distance, $\lambda$ are the control parameters representing the norm and $s$ represents the alternate cluster head choices connecting $i $ to $j$.
Further, pheromone trail updates stochastically after an ant has constructed its tour. This is done by first lowering the pheromone value on all arcs by a constant factor and then adding pheromone on the arcs the ant has crossed. The pheromone is updated in the following manner:
\begin{equation}
\tau_{ij}(t+1) =(1-p)\tau_{ij}(t)+\delta p\tau_{ij}(t+1,t)   
\end{equation}
where  $ \delta = \frac{E_{i}+E_{j}}{d^2(i,j)}$ and $p=\frac{1}{N_{res}}$ represents the pheromone evaporation rate which enables negative reinforcements where $N_{res}$ is the number of resolved nodes for minimum path to sink. The summary of ACO working is summarized  in algorithm \ref{acco}.
\begin{algorithm}[!t]
\caption{Ant Colony Optimization}\label{acco}
\begin{algorithmic} 
\STATE For each CH 
\IF {Sink Node}
	\STATE Go to next CH node
\ENDIF
\STATE Place ANTS
\STATE Search for hops
\STATE Manage generated paths
\STATE Take best path for a CH and visited nodes
\STATE Update pheromone
\end{algorithmic}
\end{algorithm}
\subsection{Distributed Intra Cluster Topology}
Within each cluster, our nodes work in a distributed manner. Event detection is primarily done at intra cluster level so in order to accommodate an infrastructure for both proactive and reactive event monitoring, a margin is defined which is used to select the  type of clustering algorithm and for handling model complexity and latency trade off.  We assume a TDMA scheduling for both inter and intra cluster routing. Within each cluster, the nodes form an undirected connected graph for extensive coverage. If $G$ denotes the given cluster at time $t$, then in order to traverse the graph, a cost factor is associated to monitor each link . For this purpose, the cost factor accommodates link quality indicator, energy and coverage. The idea is to determine trade-offs between the cost factors and react accordingly. Let there be N number of nodes in a cluster and after the choice of cluster head we need to efficiently increase linking of intra cluster nodes with cluster heads and reduce packet collisions. For this purpose, we adopt a dynamic method of graph optimization based on TDMA scheduling.
\subsubsection{Cost Function}
We introduce a cost function which is based on the link quality indicator ($L_{qi}$) at transmitter side, residual node energy and coverage diameter (based on node density and competitive radius). Here, $L_{qi}$ may be affected either by a bad wireless channel or with an actual event specially in the networks where RSSI is used as a sensor measurement. Hence, $L_{qi}$ at the transmitter is affected by residual energy and event detection. So in order to minimize coverage and maximize event detection, the cost function incorporates user defined control parameters so as to form a more scalable model depending on type of event and node as shown below:
\vspace{-0.03in}
\begin{equation}
C_{min , \mu p} = p D_{c} +(1-p)[\mu L_{qi}+ (1-\mu) E_{res}]
\end{equation}

where $p = 1/N_c$ is the reciprocal of the number of connections of each mote $N_c$, $D_c$ is the sum of the RSSI values of each connected mote, $\mu$ is defined as the tradeoff between $L_{qi}$ and $E_{res}$ and depends on transmitter parameters. Thus it is a normalized (user defined) control parameter for cost determination at the transmitter side (linked to event probability), $E_{res}$ is the residual energy at a node. 

  \subsubsection{Comparison of Routing Algorithms}
Floyd-Warshall solves all-pairs shortest path problem while Dijkstra's and Bellman-Ford algorithms are used for the single-source shortest path problem. Table \ref{tabfloyd} shows the comparison of these algorithms helping us to decide which algorithm to use. We make use of the extended version of Floyd Warshal algorithm for distributed nodes. The algorithm \ref{floyd} summarizes this extended version as it has the least complexity of all.

\begin{table*}[!t]

\centering
    \begin{tabular}{ | c | c | c | c |}
    \hline

\textbf{Sparse graph with -ive edge weights}& \textbf{Sparse graph with non -ive edge weights}&\textbf{Dense Graph} &\textbf{Dense Graph}\\ \hline
 
Johnsons Algorithm	&Dijkstras Algorithm &Floyd Warshall &Extended Floyd Warshall\\ \hline
O($n^2\log n$)	&O($n^2\log n$)&	O($n^3$)&O($n^2$)\\ \hline
    
    \end{tabular}
    \caption{Comparison of complexities of different routing algorithms}
    \label{tabfloyd}
    \end{table*} 

\begin{algorithm}[!t]
\caption{Extended Floyd Warshal Path Reconstruction }\label{floyd}
\begin{algorithmic} 

\STATE Record the visited nodes
\STATE Crawl up to head node
\IF  {multiple node requests}
	\STATE select the branch with maximum nodes linked
\ELSE
\STATE Pass the branch
\ENDIF
\end{algorithmic}
\end{algorithm}

\subsection{Intra-cluster Design}
\begin{figure}[!t] 
	\centering
	\includegraphics[width=0.38\textwidth]{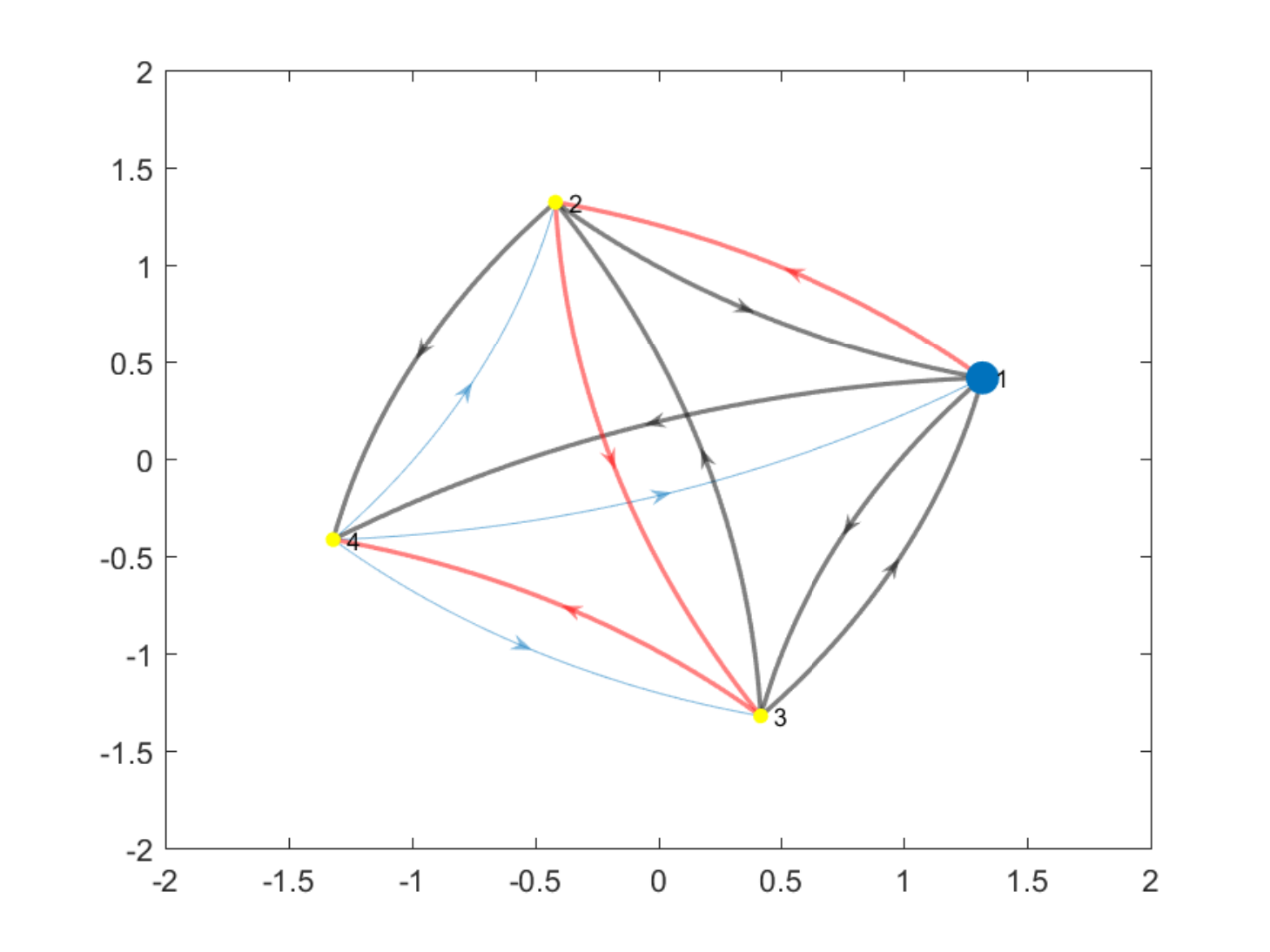}
	 \caption{\footnotesize Representation of distributed intra cluster routing with 4 nodes; the red branch represents the path with minimum cost and maximum
node traversal}
	 \label{ICD}
\end{figure}

The resultant final algorithm for intra cluster distributed routing is summarized in algorithm \ref{ICDR}. The implementation of distributed intra cluster routing  with 4 nodes is demonstrated in the Figure \ref{ICD}. Each node in distributed cluster forms a branch (multiple links) from different nodes to CH. The CH selects a branch with minimum cost and maximum node traversal as indicated by the red branch in the Figure \ref{ICD}.  Since it is a strongly connected graph, there is link from each node to other node, so in order to improve the network flow, we only need to select one branch at a time. In this way for reducing the overhead in communication, we dig up an optimal dynamic route from connecting all the nodes in the cluster and communicate the information to the CH. 
The resultant final algorithm for intra cluster distributed routing is summarized in algorithm \ref{ICDR}. The implementation of distributed intra cluster routing  with 4 nodes is demonstrated in the Figure \ref{ICD}. Each node in distributed cluster forms a branch (multiple links) from different nodes to CH. The CH selects a branch with minimum cost and maximum node traversal as indicated by the red branch in the Figure \ref{ICD}.  Since it is a strongly connected graph, there is link from each node to other node, so in order to improve the network flow, we only need to select one branch at a time. In this way for reducing the overhead in communication, we dig up an optimal dynamic route from connecting all the nodes in the cluster and communicate the information to the CH. 
\begin{algorithm}[!t]
\caption{Intra Cluster Energy Efficient Distributed Routing and Monitoring}\label{ICDR}
\begin{algorithmic} 
\STATE For all nodes in the Cluster
\STATE Broadcast link to all neighboring nodes
\STATE  Calculate cost function and choose path of minimum cost
\IF {hard margin}
	\STATE Do Reactive monitoring
\ELSE 

\STATE Do Proactive monitoring
\ENDIF
\STATE Do extended Floyd-Warshall Optimization
\end{algorithmic}
\end{algorithm}

\subsection{Modified Deep belief network for hybrid event monitoring}
\begin{figure}[!t] 
	\centering
	\includegraphics[width=0.49\textwidth]{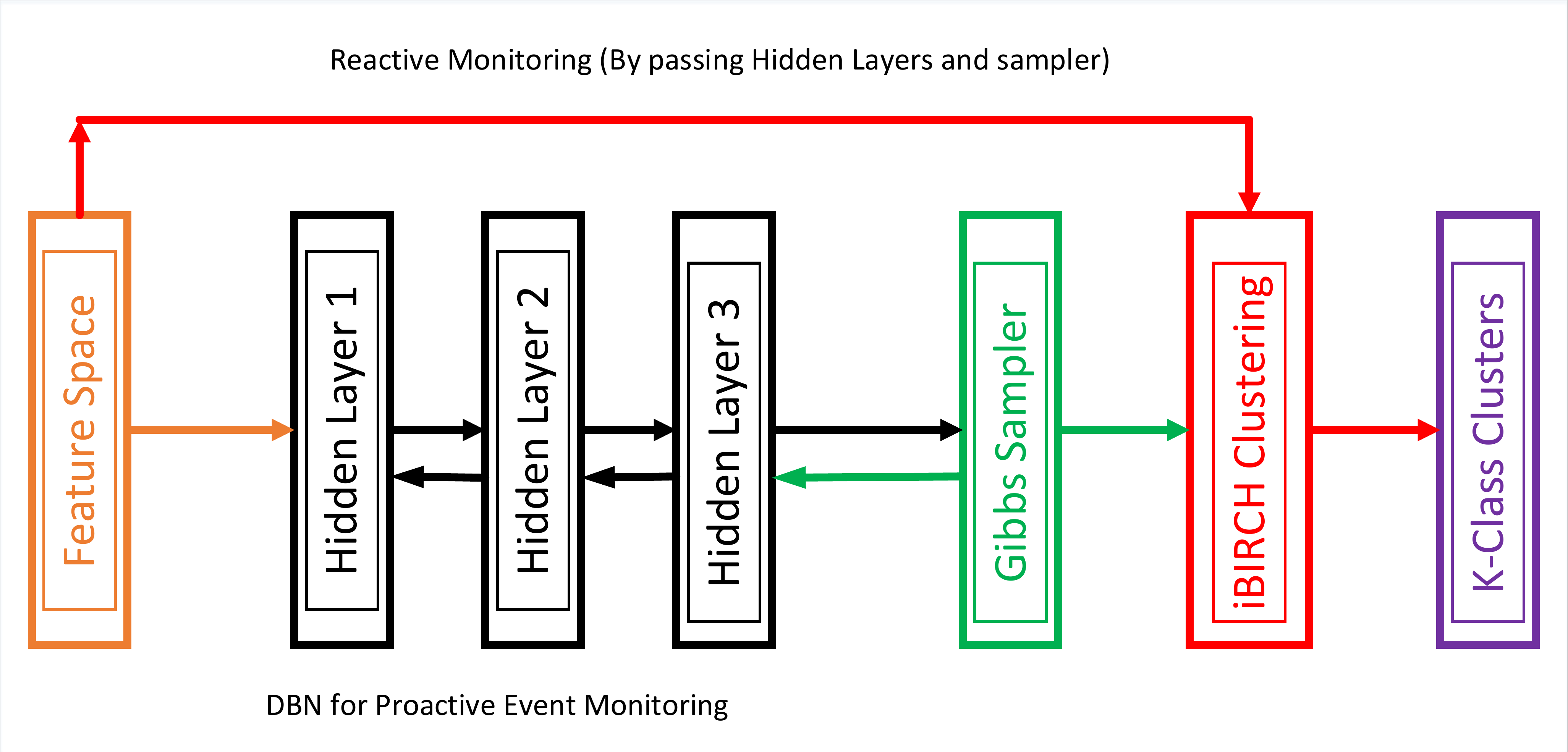}
	\caption{\footnotesize The modified form of the Deep Belief Network with three hidden layer; these layers are bypassed in case of reactive monitoring}
	\label{mod_dbn}
\end{figure} 

The event monitoring needs to be flexible to accommodate instances with both proactive and reactive event monitoring. For this purpose, a user defined margin is integrated within the event clustering algorithm so that we can switch between two degrees of freedom. But using two different clustering techniques within the same framework increases the visit time (data aggregation and processing time at a node). Therefore, we make use of a modified implementation of \emph{Deep Belief Network}   [\ref{leachEEE19}], [\ref{leachEEE21}], [\ref{leachEEE22}] with 3 hidden layers and online Gibbs sampling for proactive monitoring. All of this would be bypassed in case of user defined event thresholds are crossed as shown in the Figure \ref{mod_dbn}.

In order to minimize noise data, a stochastic framework to reconstruct inputs from unsupervised data has been used which makes event decision  utilizing incremental balanced iterative reducing cluster hierarchy (iBIRCH). In our implementation, Stacked Restricted Boltzmann Machines accompanied with Gibbs sampling and iBirch are used for clustering in generative deep belief network. Maximum likelihood is derived from data values and is passed for prior selection after bottom up pass, sampling and bottom down pass to shape the posterior distribution and selection of hidden nodes. A layer is learnt greedily before fine tuning of parameter in above layer. After fine tuning, the parameters in lower layer are set by back propagation algorithm.
We have developed Gaussian-Bernoulli Restricted Boltzman Machines (RBM) based DBN with visible real-valued RSSI values $v \in {R^D}$ and transformed $h \in{\{0,1\}}  ^F$ as stochastic binary hidden variables. The energy of the joint state $\{v, h\}$ of the Gaussian RBM is defined as:
  \begin{equation}
 E(v,h; \theta) = \sum_{i=1}^{D} \frac{(v_{i}-b{i})^2}{2 \sigma_{i}^{2}} - \sum_{i=1}^{D} \sum_{j=1}^{F}W_{ij}h_{j} \frac{v_{i}}{\sigma_{i}} -\sum_{j=1}^{F}a_{j}h_{j}  
 \end{equation}
 
 where $\theta=\{W,a,b,\sigma^2\}$ are the model parameters. $W_{ij}$ represents the symmetric interaction term between visible variable $v_{i}$ and hidden variable $h_{j}$,and $b_{i}$ and $a_{j}$ are bias terms.
 The joint distribution over the visible and hidden variables is defined by:
 \begin{equation}
 P(v,h; \theta) =  \frac{1}{Z(\theta)}\exp(-E(v,h:\theta))
 \end{equation}
 where $Z(\theta)$ is known as the partition function or normalizing constant. The model then assigns the marginal distribution over the visible vector $v$, which takes the following form:
 \begin{equation}
 P(v; \theta) = \sum_{h} \frac{\exp(-E(v,h;\theta))}{ \int_{v'} \sum_{h} \exp(-E(v,h;\theta)) dv'}
 \end{equation}
 From Equation 7, derivation of the following conditional distributions is straightforward:
 \begin{equation}
p(v_{i}=x|h)=\frac{1}{\sqrt{2 \pi \sigma^2}} \exp\left(\frac{-(x-b_{i}-\sigma_{i} \sum_{j} h_{j}W_{ij})^2}{2 \sigma_{i}^2}\right)
\end{equation}
\begin{equation}
p(h_{j}=1|v)= g(h_{j})+ \sum_{i}W_{ij} \frac{v_{i}}{\sigma_{i}}
\end{equation}
where g(x)=$\frac{1}{1+\exp(-x)}$ is the logistic function.
Each visible unit is modeled by a Gaussian distribution whose mean is shifted by the weighted combination of the hidden unit activations. Given a set of observations $\{v_{n}\}_{n=1}^N$, the derivative of the log-likelihood with respect to the model parameters $W_{ij}$ takes the following form: 
 \begin{equation}
\frac{\partial \log P(v;\theta)}{\partial W_{ij}}=E_{P_{data}}\left[\frac{v_{i}h_{j}}{\sigma_{i}}\right]-E_{P_{model}}\left[\frac{v_{i}h_{j}}{\sigma_{i}}\right]
\end{equation}
where $E_{P_{data}}[.]$ denotes an expectation with respect to the data distribution $P_{data}(h,v;\theta) = P(h|v;\theta)P_{data}(v)$, where $P_{data}(v) = \sum_{n}\delta(v-v_n)$ represents the empirical distribution, and $E_{P_{model}}[.]$ is an expectation with respect to the distribution defined by the model, as in Equation 8.
  
Exact maximum likelihood learning in this model is intractable because exact computation of the expectation $E_{P_{model}}[.]$ takes time that is exponential in $\min(D, F)$, i.e., the number of visible or hidden variables. So, learning is done by following an approximation to the gradient of a different objective function, called the Contrastive Divergence (CD) algorithm. 

\begin{equation}
\Delta W = \alpha (E_{P_{data}[vh^T]}- E_{p_{T}}[vh^T])
\end{equation}
where $\alpha$ is the learning rate and $P_{T}$ represents a distribution defined by running a Gibbs chain initialized at the data for T full steps. The special bi-partite structure of RBMs allows for an efficient Gibbs sampler that alternates between sampling the states of the hidden variables and giving the states of the visible variables independently. The DBN model here uses an incremental Gibbs sampler, an extension of [\ref{leachEEE21}] where we use a set of topics (features dimensioned to fit in the hidden layers) from the feature space of sensor outputs. The conditional PDF of the topic variables is given as:

  \begin{equation}
 P(z_{j}|z_{i},w_{i}) \propto n_{z_{ij}}^{w_{j} }+ \Theta n_{z_{ij}}^{d_{j}}+ \gamma
 \end{equation}
where $\Theta$ and $\gamma$ are tuning parameters, $\{z_1,. ., z_{j-1}, z_{j+1},. ., z_N \}$ are the samples of the the topic variables. The topic assignment of word $j$ is sampled according to its conditional distribution given above where $w_i$ is the size of the  feature space $i$, $n_{z_{ij}}^{w_{j}}$ is the number of times feature $w_j$ is assigned to topic $z_j$ and  $n_{z_{ij}}^{d_{j}}$ is the number of times a row entry $d_j$ is assigned to topic $z_j$.  


\section{Evaluation}
The proposed optimized sensor network has been simulated in MATLAB. The rounds (the time lapse between the request generated by sink node and the data received by all CH i.e., complete round trips) have been normalized and are fixed  for the evaluation of proposed and existing designs. The formal and proposed topologies have been given same initial length and width of the network area, same initial energy of each node, same energy required for transmission of each bit and same energy required for reception of each bit.

\subsection{Inter-cluster Optimization}
\begin{figure}[!t] 
	\centering
	\includegraphics[width=0.38\textwidth]{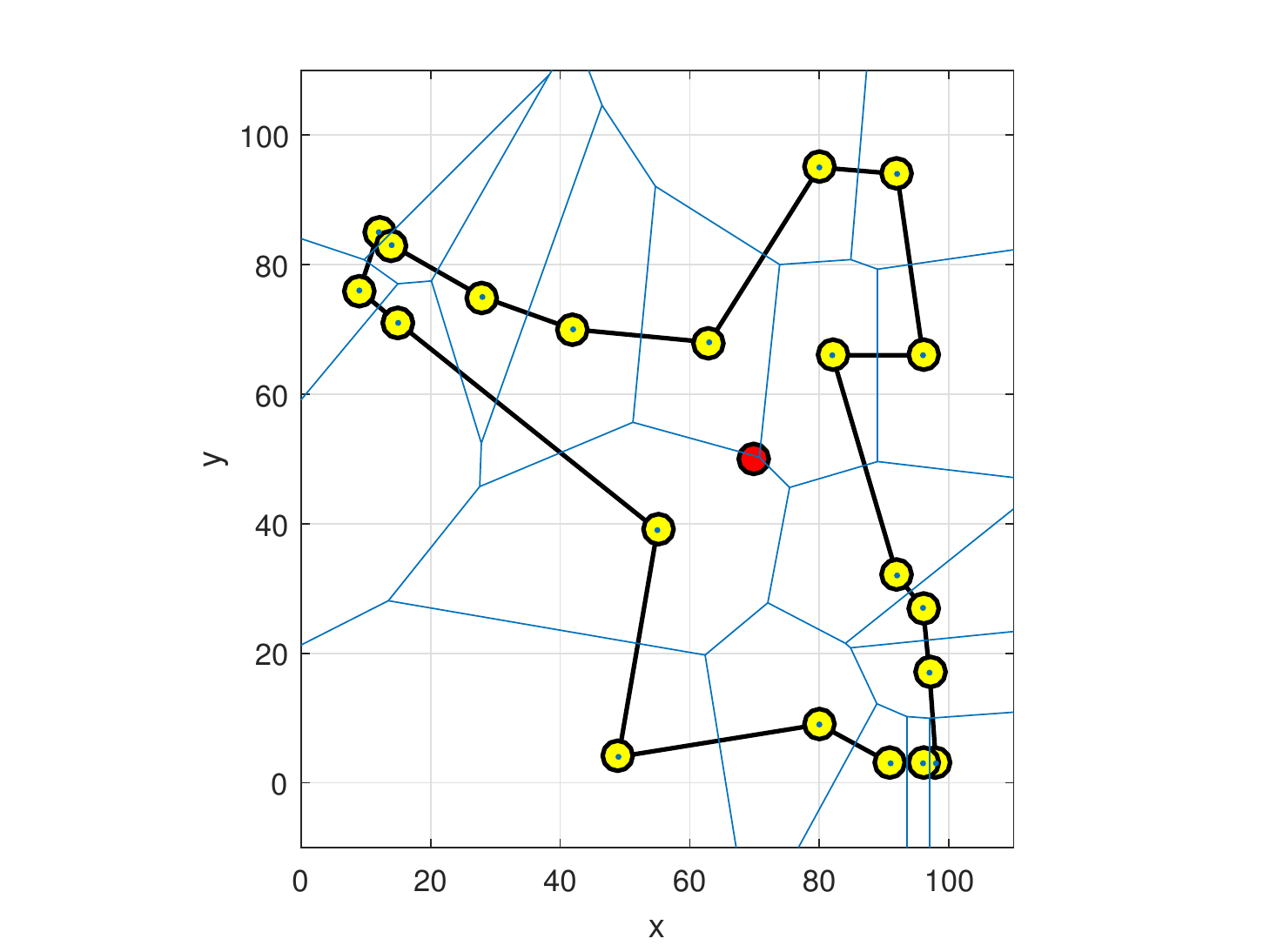}
	\caption{Simulation of inter cluster optimization}
	\label{ICO}
\end{figure}

In order to check our method  of inter cluster optimization, we implemented a network of 200 nodes from which 20 nodes were selected as cluster heads. The resultant coverage based on Voronoi tessellation depending on node density is shown in Figure \ref{ICO}. Higher the node density, more will be the coverage area of the cluster. After setting up clusters, ACO is used for optimal routing among the CH to decide between multi hop and parallel data retrieval by sink node.

In Figure \ref{ICO}, yellow dots indicate the cluster heads (CH) and the red dot indicates the sink node.
As a result of ACO, cost convergence indicates optimal routing of cluster heads, which calculates energy cost function minimization in 100 iterations at sink node. At a system level, the performance metric used is number of packets sent to the sink node, the number of dead nodes and energy decay versus number of iterations or rounds. We have compared our algorithm implementation with LEACH and LEACH-EEE as shown in Figures 5 and 6.
 
\begin{figure*}
  \centering
	\includegraphics[width= 1 \linewidth,height=2in]{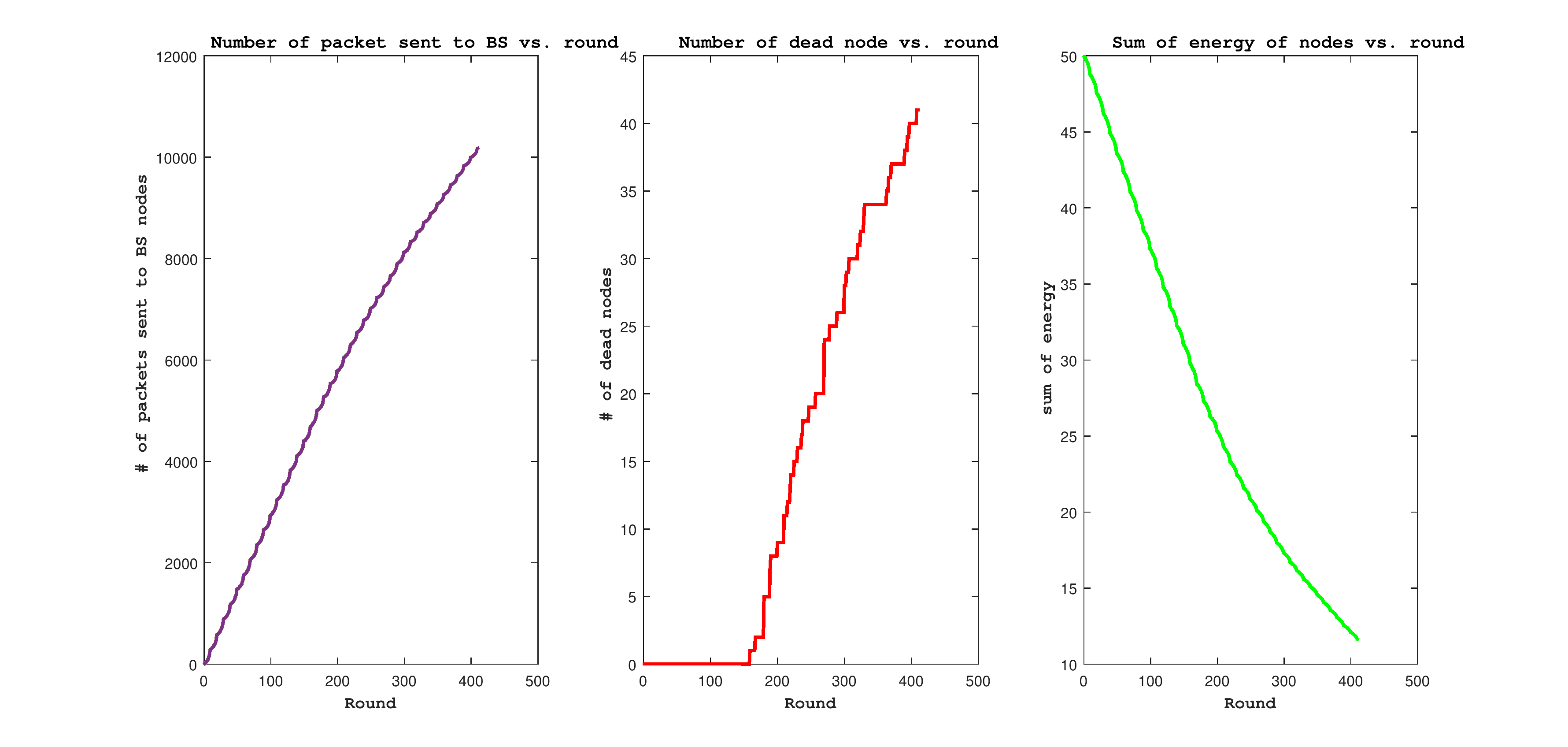}\par
  \caption{\footnotesize The performance with the proposed Optimized LEACH vs iteration or rounds (a) Number of transmitted packets (b) Number of Dead Nodes (c) Sum of energy of nodes}
\label{opleach}
  \vspace*{\floatsep}

\end{figure*}
\begin{figure*}
 \centering
\includegraphics[width=1\linewidth,height=2in]{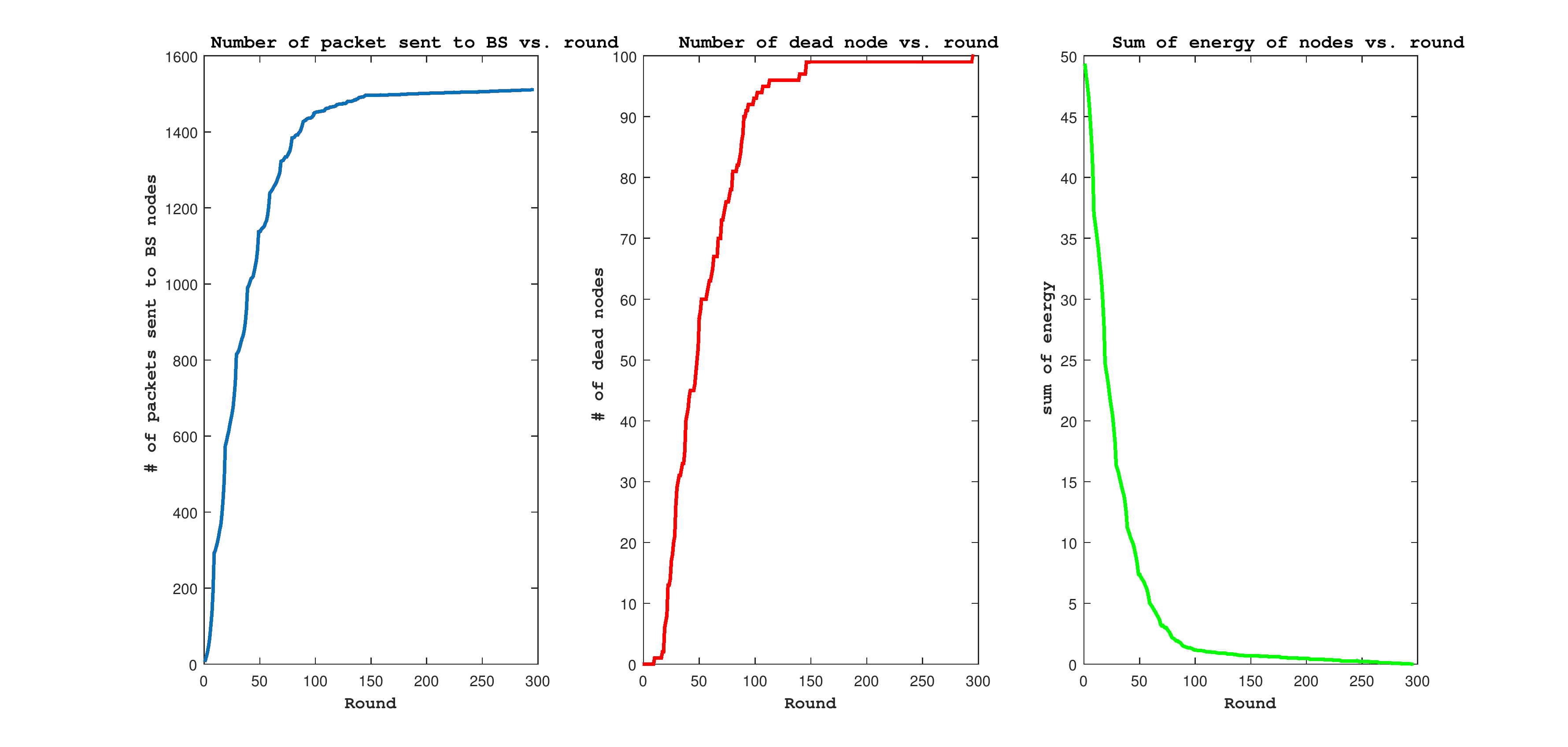}\par
  \caption{\footnotesize The performance with LEACH-EEE}
\label{leacheee}
  \vspace*{\floatsep}

\end{figure*}%

\begin{table}[!t]
\centering
	\begin{tabular}{ | c | c | c | }
	\hline

	\textbf{Hidden Units per layer}& \textbf{Total Units}&\textbf{Error rate (\%)}\\ \hline
	 
	12-12-12	&36	&2.78\\ \hline
	20-15-12	&47&	2.33\\ \hline
	20-20-15&	55	& 2.13\\ \hline
	20-20-20	&60	&2.45\\ \hline
			
	\end{tabular}
\caption{\footnotesize Performance improvement with different units of hidden layers in DBN; the error rate gets significantly reduced with the proposed strategy}
\label{DBN results}
\end{table} 

Figure \ref{opleach} and \ref{leacheee}  show the comparison of the results of our optimized method and state of the art LEACH EEE protocol. Figure \ref{opleach}(a) shows that our model has a throughput of almost 6 times greater than LEACH-EEE and has notably higher throughput over a longer network lifetime. Figure \ref{opleach}(b) shows that the first dead node appears after approximately 160 rounds whereas in each LEACH-EEE, a mote is dead (energy below threshold) appears after only 10 rounds (see Figure \ref{leacheee}(b)). Note that the quantitative numbers here are normalized for comparison to give an overall network behavior. It can also be seen from Figure \ref{opleach}(c) and \ref{leacheee}(c) that the rate at which the total energy of the overall network drains is lower for our optimized method than LEACH-EEE. The energy drainage rate of LEACH-EEE is approximately ten times to that of proposed LEACH EEE. Note that we assume the network to be dead when half of the motes in the network are dead (100 in this case). Here, it is important to note that staircase decay of nodes indicates that though the overall total energy of the system is going to decay i.e., dead nodes will increase with time but the energy distribution is randomized and is ``managed optimally'' i.e., the stairs in the graph represents the optimal management of total system energy (as in Figure \ref{opleach}(a)).
\subsection{Intra Cluster Optimization}
For intra cluster optimization, we investigated  different partitioned clustering algorithms on the obtained data including EM, Farthest First, Density based, OPTICS and simple K-means, however agglomerative iBRICH algorithm showed higher performance once trained.  This reactive monitoring is  event triggered. For more complex proactive monitoring of multiple physical factors, a proactive deep belief network is used which further reduced the error rate on test data as shown in the Table \ref{DBN results}. The results show a performance improvement with different units of hidden layers in DBN; the error rate gets significantly reduced with the proposed strategy and lies in the range of 2.1 to 2.8 whereas with only iBRICH clustering, the error rate is around 5\%. 
\section{Conclusion}
In this paper, we propose a modified LEACH algorithm which minimizes the overall energy consumption of the network and at the same time achieves significantly better throughput. The proposed solution forms optimized cluster boundaries using Voronoi tessellations. Thus dynamic cluster boundaries are formed on the basis of node density and energy consumptions. Voronoi Tessellations allow us to minimize energy and maximize the radius of each cluster. Once the cluster boundaries have been finalized, Ant Colony Optimization is used to determine the path with minimum cost from the CHs to the sink in a multi-hop communication model. With in a cluster, dynamic optimization is used to minimize the cost. The result show that the proposed optimized LEACH has a throughput of almost 6 times greater than the state of the art and that too with much better network life time (an order of magnitude better).

\end{document}